\documentclass[reprint, superscriptaddress, amsmath, amssymb, aps,showpacs]{revtex4-1}

\usepackage{graphicx}
\usepackage{dcolumn}% Align table columns on decimal point
\usepackage{bm}% bold math
\usepackage{mathtools}

\def\sup{|\hspace{-0.5mm}\uparrow\hspace{0.5mm}\rangle}
\def\sdown{|\hspace{-0.5mm}\downarrow\hspace{0.5mm}\rangle}
\def\csup{\langle\hspace{0.5mm}\uparrow\hspace{-0.5mm}|}
\def\csdown{\langle\hspace{0.5mm}\downarrow\hspace{-0.5mm}|}

\begin{document}

\title{Quasienergies and dynamics of superconducting qubit in time-modulated field}% Force line breaks with \\

\author{Gor A. Abovyan}
\email{gor.abovyan@ysu.am}
\affiliation{Yerevan State University, A. Manookyan 1, 0025, Yerevan, Armenia}
\affiliation{Institute for Physical Research, Ashtarak-2, 0203, Ashtarak, Armenia}

\author{Gagik Yu. Kryuchkyan}
\email{kryuchkyan@ysu.am}
\affiliation{Yerevan State University, A. Manookyan 1, 0025, Yerevan, Armenia}
\affiliation{Institute for Physical Research, Ashtarak-2, 0203, Ashtarak, Armenia}

\date{\today}

\begin{abstract}

We analyze the dynamics of a superconducting qubit and the phenomenon of multiorder Rabi oscillations in the presence of a time-modulated external field. Such a field is also presented as a bichromatic field consisting of two spectral components, which are symmetrically detuned from the qubit resonance frequency. This approach leads to obtaining qualitative quantum effects beyond those for the case of monochromatic excitation of qubits. We calculate Floquet states and quasienergies of the composite system ''superconducting qubit plus time-modulated field" for various resonant regimes. We analyze the dependence of quasienergies from the amplitude of an external field, demonstrating the zeros of difference between quasienergies. We show that, as a rule, populations of qubit states exhibit aperiodic oscillations, but we demonstrate the specific important regimes in which dynamics of populations becomes periodically regular.

\end{abstract}

% PACS, the Physics and Astronomy Classification Scheme.
\maketitle

%\tableofcontents

\section{\label{Introduction}Introduction}

Superconducting circuits based on Josephson junctions are promising candidates for studying fundamental physics and implementing qubits and controllable quantum two-level systems for quantum computing (see, for example, \cite{Makh,You,Clarke,Scho} for reviews).
The simplest Josephson-junction (JJ) qubit consists of a small superconducting island with $n$ excess Cooper-pair charges connected by a tunnel junction with capacitance $C_J$ and Josephson coupling energy $E_J$ to a superconducting electrode and the single-electron charging energy $E_C$. In the case of a qubit only two charge states with $n=0$ and $1$ play a role while all other charge states, having a much higher energy, can be ignored. Thus, a superconducting charge qubit \cite{Nak} behaves as an artificial two-level atom in a Cooper box, which is well described by two charge states, and the electrostatic energy difference between these states is controlled by the normalized gate charge. 

When a qubit is driven by an external periodically time-dependent electromagnetic field, it has given rise to new quantum effects such as Rabi oscillations and coherent control \cite{Vion,Johan,Naka,Yu}, which are the bases for quantum operations. In a series of experiments many fundamental effects from quantum optics have been demonstrated \cite{Schus,Frag,Astaf,Neel,Astaf1,Astaf2,Abdu}, including a lasing effect with a Josephson-junction charge qubit embedded in a superconducting resonator \cite{Astaf}. Superconducting qubits usually have short coherence time; therefore, to decrease the time for performing gate operations a large-amplitude external field should be applied. The dynamics of a qubit driven by large-amplitude external fields in the case of driving around the region of avoided level crossing has been also studied (see, \cite{Ash} and \cite{Shev} for reviews).

Most studies of qubit dynamics assume the driving field to be monochromatic or a single cavity mode. In the present paper we investigate dynamics of a qubit and the phenomenon of Rabi oscillations for an artificial two-level atom interacting with a monochromatic field with time-modulated amplitude. Such an external field can be also presented as a bichromatic field that consists of two components of equal amplitudes which are symmetrically detuned from the qubit resonance frequency. In this case, the modulation frequency is displayed as the difference between frequencies of two spectral components. 
This approach, involving modulation of the energy splitting of a qubit in complicated form due to interaction with an external bichromatic field, is different from the standard scheme of laser physics in which the bichromatic field leads to dipole transitions between two states of atoms.  
This approach can be also applied for investigation of a wide variety of interesting phenomena including tunneling dynamics of time-dependently driven nonlinear quantum systems. In addition, this problem offers an ideal testing ground for studying the fundamental interactions between qubits and multi-spectral component light. Note, that the  scheme of the Josephson-junction qubit considered in this paper seems to be close to the  experimental scheme on the frequency-modulated transmon qubit performed most recently in Ref. \cite{Paraoanu}. 

The other goal of this paper is application of the method of quasienergies and quasienergetic states (QESs) (or the so-called Floquet states) for the qubit in a bichromatic field. Note that, at first, the QESs of the composite system consisting of an atom and time-periodic e.m. field have been considered in \cite{Shirley,Zeldovitch,*ZeldovitchZh,Ritus,*RitusZh}. These states provide a classical counterpart to well-known atomic-dressed states \cite{Cohen} in which the coupling to the laser is described by a classical field, whereas the coupling to the vacuum must be described in second quantization. However, one may still hope that in the limit of a macroscopically relevant laser field, both approaches lead to the same results. On the other hand, a certain advantage of the classical treatment implied by the Floquet approach lies in the fact that laser pulses can be handled more easily than in a fully quantized approach to the field (see, e.g., \cite{Glusko}). In the Floquet picture the QESs of the composite system are formed in a strong external field, and the radiation processes and spectral lines are described by transitions between them due to the interaction of the composite system with an electromagnetic vacuum or with a weak probe field. In this way, the master equations in the QES basis were obtained in \cite{KryuchkyanTeorFiz,*KryuchkyanJEPT,Kryuchkov}
and in the dressed-state basis in Ref. \cite{Cohen1}. Thus, the method of QESs is a powerful theoretical framework for the study of bound-bound multiphoton transitions driven by periodically time-dependent fields (see, for review \cite{Shih-I}). There have been several experiments on nonlinear and quantum optics that have been interpreted in terms of quasienergy levels including basic experiments on the resonance fluorescence and the probe absorption spectroscopy for a two-level atom in a strong laser field.
QESs and dressed states have also been used in areas of radiation corrections to atomic levels in the presence of a strong laser field, including the calculation of the Lamb shift \cite{KryuchkovTeorFiz1982,*KryuchkovJEPT1982,KryuchkyanIzvestija,*KryuchkyanIzvestija2,JentEverHaas2003,EverJentKeit2004,JentKeit2004,KryuJentEverKeit2007}.
The dressed-state approach including atomic motion was introduced in Ref. \cite{CohenDalib1985}, while the QES method was used for strongly confined ions in Refs. \cite{KryuchkyanOC1998,JakKneeKry1998} for multiphoton processes with laser-cooled and trapped ions, for the scheme of an ion-trap laser \cite{KryKee1999Lett,KryKee1999Rev}, and for investigations of photons correlation in an ion-trap system \cite{JakobKryuchkyan1999Rev}.

Applications of QESs and quasienergies to Josephson qubits in a driving field have been done in several papers \cite{Shev,Russomanno,Silveri,Tuorila}, including a review paper on Landau-Zener-St{\"u}ckelberg interferometry \cite{Shev}, probe spectroscopy of QESs \cite{Silveri}, application of the Floquet theory to Cooper pair pumping \cite{Russomanno}, and observation of the Stark effect and generalized Bloch-Siegert shift in the experiment with a superconducting qubit probed by resonant absorption via a cavity \cite{Tuorila}. The experiments on the Rabi oscillations in monochromatically driven Josephson qubits have been performed and interpreted on the basis of dressed states \cite{Johan,Naka}. 

QESs for a two-level atom in the bichromatic field have also been studied in a series of papers (see, for example, \cite{Kryuchkov, Freed1,*Freed2,Jakob1,Jakob2,Jakob3}). 
Note that investigations of bichromatically driven natural two-level systems have a long history in areas of laser physics, nonlinear optics, and quantum optics. The corresponding Hamiltonians of such systems involve the coupling of a bichromatic field to the transition dipole moment between two states of atoms in contrast to the case of a superconducting qubit in a Cooper box in which an external field only drives the atomic energetic levels. The spectrum of resonance fluorescence (RF) of a two-level atom in a bichromatic field was calculated in \cite{Kryuchkov,Freed1,Freed2,Tew}. The fluorescence spectrum for the general case of arbitrary detuning was obtained in Refs. \cite{Zhu,Ficek} and was observed experimentally in Ref. \cite{Agarwal} in agreement with the theoretical results. Effects of cavity-modified dynamics were also found in Ref. \cite{Agarwal1} for two-level Rydberg atoms in a microwave cavity under the influence of a bichromatic field. In a series of papers it has been demonstrated that photon correlation and quadrature squeezing induced by a bichromatic field are drastically different from the case of RF in a monochromatic field \cite{Jakob1,Jakob4,Jakob2,Jakob3}. We especially focus on unusually strong superbunching effects in the second-order correlation function as a result of strongly correlated two photon emissions at the frequency of atomic transition \cite{Jakob1,Jakob2} in applications for two-photon lasing. 

We believe that the results of forming atomic spectral lines with strongly different frequencies under bichromatic radiation are important also for the superconducting qubit inducing additional Rabi oscillations on quasienergetic states of the qubit. Additionally, we demonstrate below that quasienergetic states and quasienergies of the bichromatically driven superconducting qubit under consideration differ drastically from the analogous well-known states of the standard two-level atom in a bichromatic field, and due to this difference unusual field-dependence effects appear for the qubit.

Note, that time modulation of quantum dynamics for some systems allows effective control of dissipation and decoherence effects, essentially improving the quantum effects. Indeed, it has been shown that the time modulation in an optical parametric oscillator leads to improvement of squeezing and continuous-variable entanglement of generated modes \cite{Adam1,Adam2}, and application of such an approach to an anharmonic oscillator leads to preparation of oscillatory Fock states' superpositions in the presence of decoherence \cite{Gevorg1,Gevorg2}. Thus, we expect that this approach applied to artificial atoms, particularly superconducting qubits, will lead to obtaining new qualitative quantum effects involving control of superconducting qubits and improvement of decoherence.  Nevertheless, in this paper, as the first part of these investigations, we only consider nondissipative dynamics of a qubit in a time-modulated field for short time intervals. 

In this paper, we present analytical results for nontrivial dynamics of a qubit in a time-modulated field (a bichromatic field), particularly, considering in detail time-dependent populations of qubit states. We calculate QESs and quasienergies of the composite system ''superconducting qubit plus time-modulated field" in resonance approximation by using the Furry picture. 

The paper is arranged as follows. In Sec. II we derive the Hamiltonian of the system in the resonance approximation. In Sec. III we consider the tunneling amplitude of transitions between states of a qubit in the presence of a time-modulated (or bichromatic) field and calculate corresponding QESs as well as quasienergies. Then, in Sec. IV we investigate the properties of quasienergies as well as compare two systems with different time-dependent components along $x$ and $z$ axes. In Sec. V the Rabi oscillations physics of the qubit driven by a time-modulated field is considered. We summarize our results in Sec. VI.

\section{Furry picture for qubit in time-modulated field}

The qubit is realized if the charging energy of a superconducting electron box is much larger than the Josephson coupling energy. In the regime of low-level excitation the system is formed by two charge states: $\sdown$ and $\sup$ which have either zero Cooper pairs or one Cooper pair. Thus, the system that we consider here is a qubit coupled to a time-modulated field (or a bichromatic field) with the Hamiltonian
\begin{equation}\label{Ham}
\hat{H}(t)=\hat{H}_{0}+\hat{H}_{V},
\end{equation}
where
\begin{equation}\label{HamParts}
\hat{H}_{0}=-\frac{1}{2}(\varepsilon_{0}+f(t))\hat{\sigma}_{z},\hspace{0.5cm} \hat{H}_{V}=-\frac{\Delta}{2}\hat{\sigma}_{x}.
\end{equation}
Here, the external field reads
\begin{equation}
f(t)=2A\cos\left(\omega_{0}t\right)\cos\left(\delta t\right),
\end{equation}
where $\omega_{0}$ and $\delta$ are the central and modulation frequencies, provided that $\delta<<\omega_{0}$. 
This external field can be presented as a bichromatic field of the form
\begin{equation}
f(t) =A[\cos(\omega_{1}t)+\cos(\omega_{2}t)]
\end{equation}
with equal amplitudes of two spectral components at the frequencies $\omega_{1}=\omega_{0}-\delta$ and $\omega_{2}=\omega_{0}+\delta$.

Here, $\varepsilon_{0}=E_{Q}(1-2n_{g})$ is the electronic energy difference between the ground and excited states of the qubit and $\Delta=E_{j}$ is the Josephson coupling energy or the tunneling amplitude between the basis states. The operators $\hat{\sigma}_{x}$, $\hat{\sigma}_{z}$ denote the Pauli spin matrices: $\hat{\sigma}_{z}=\sup\csup-\sdown\csdown$, $\sigma_{x}=\sup\csdown+\sdown\csup$. The Hamiltonian Eq. (\ref{Ham}) describes various physical systems in addition to the JJ artificial atom \cite{Hanggi}. In general, it describes the tunneling dynamics of bichromatically driven nonlinear quantum two-level systems. 

It should be noted that very often in area ''atom+laser" interaction the other Hamiltonian is used, in which the coupling of a time-dependent electromagnetic field to the transition dipole moment between two states of atoms takes place in contrast to the case of a superconducting qubit, where an external field drives the atomic energetic levels Eqs. (\ref{Ham}) and (\ref{HamParts}). The corresponding Hamiltonian $\hat{H}_{at}$ describing interaction along the $x$ axis can be related to the Hamiltonian Eq. (\ref{Ham}) with the time-dependent component along the $z$ axis by a rotation around the $y$ axis. The result reads 
\begin{equation}\label{AtomHam}
\hat{H}_{at}=e^{-i\frac{\pi}{4}\hat{\sigma}_{y}}\hat{H}(t)e^{i\frac{\pi}{4}\hat{\sigma}_{y}}=-\frac{1 }{2}\Delta\hat{\sigma}_{z}-\frac{1}{2}(\varepsilon_{0}+f(t))\hat{\sigma}_{x}.
\end{equation}
The later Hamiltonian is typical for a natural two-level atom interacting with a bichromatic field. In this case, the parameter $\Delta$ describes an energy difference and the interaction term is responsible for the transitions between two atomic states. See also Sec. IV.B.

We describe the dynamics of the system in the Furry-state representation $|\Psi(t)\rangle=U(t)|\Psi_{U}(t)\rangle$, in which the equation for the vector state of the full system is
\begin{equation}\label{Shred}
i\frac{\partial}{\partial t}|\Psi_{U}(t)\rangle=\hat{H}_{I}|\Psi_{U}(t)\rangle.
\end{equation}
The interaction Hamiltonian is given by
\begin{equation}\label{HamInt}
\hat{H}_{I}(t)=\hat{U}^{-1}(t)\hat{H}_{V}\hat{U}(t)=-\frac{\Delta}{2}\hat{U}^{-1}(t)\hat{\sigma}_{x}\hat{U}(t)
\end{equation}
while the unitary operator $\hat{U}(t)$ obeys the equation of motion
\begin{equation}
i\frac{\partial}{\partial t}\hat{U}(t)=\hat{H}_{0}\hat{U}(t).
\end{equation}
It is easy to realize that operator $U(t)$ has a simple form
\begin{equation}
\hat{U}(t)=\exp\left[-i\int^{t}_{0}\hat{H}_{0}(t')dt'\right]=\exp(i\varphi(t)\hat{\sigma}_{z}),
\end{equation}
where
\begin{equation}
\varphi(t)=\frac{1}{2}\left[\varepsilon_{0}t+\frac{A}{\omega_{1}}\sin(\omega_{1}t)+\frac{A}{\omega_{2}}\sin(\omega_{2}t)\right].
\end{equation}
Thus, the interaction Hamiltonian is calculated in the following form
\begin{eqnarray}
\hat{H}_{I}(t)=-\frac{\Delta}{2}e^{-i\varphi(t)\hat{\sigma}_{z}}\hat{\sigma}_{x}e^{i\varphi(t)\hat{\sigma}_{z}}\nonumber\\
=-\frac{\Delta}{2}\left(\begin{array}{cc}
0 & e^{-2i\varphi(t)} \\
e^{2i\varphi(t)} & 0 \end{array} \right).
\end{eqnarray}
In the $\sigma$-matrix form this Hamiltonian can be written as
\begin{equation}
H_{I}(t)=-\frac{\Delta}{2}\left[\sigma_{+}e^{-2i\varphi(t)}+\sigma_{-}e^{2i\varphi(t)}\right].
\end{equation}
For simplification of the Hamiltonian we use the following formulas with the Bessel functions 
\begin{equation}
\exp\bigg[i\frac{A}{\omega_{1}}\sin(\omega_{1}t)\bigg]=\sum_{n_{1}}J_{n_{1}}\bigg(\frac{A}{\omega_{0}+\delta}\bigg)e^{in_{1}(\omega_{0}+\delta)t},
\end{equation}
where $J_{n}(x)$ is $n$-th order Bessel function of the first kind. In the result we can obtain
\begin{eqnarray}
& & e^{2i\varphi(t)}= \sum_{n_{1}}\sum_{n_{2}}J_{n_{1}}\bigg(\frac{A}{\omega_{0}+\delta}\bigg)J_{n_{2}}\bigg(\frac{A}{\omega_{0}-\delta}\bigg)\nonumber\\
& \times & e^{i[\varepsilon_{0}+(n_{1}+n_{2})\omega_{0}+(n_{1}-n_{2})\delta]t}. 
\end{eqnarray}
We also add that $e^{-2i\varphi}=(e^{2i\varphi})^{*}$. 

The resonance condition is formulated using the requirement that the oscillating terms in time have vanished. Thus, this condition is formulated for the central frequency $\omega_{0}$  and the electronic energy difference as $\varepsilon_{0}-N\omega_{0}=\Delta_{N}<<\varepsilon_{0}$, where $n_{1}+n_{2}=-N$. In this approximation we obtain 
\begin{eqnarray}
e^{2i\varphi} & = & e^{i\Delta_{N}t}\sum_{n_{1}+n_{2}=-N}J_{n_{1}}(z_{1})J_{n_{2}}(z_{2})e^{i(n_{1}-n_{2})\delta t}\nonumber\\
& = & e^{i\Delta_{N}t}\sum_{n_{1}}J_{n_{1}}(z_{1})J_{-N-n_{1}}(z_{2})e^{i(2n_{1}+N)\delta t}\nonumber\\
& = & e^{i(\Delta_{N}+N\delta)t-iN\pi}\sum_{n_{1}}J_{n_{1}}(z_{1})J_{N+n_{1}}(z_{2})e^{in_{1}\gamma},
\end{eqnarray}
where $z_{1}=\frac{A}{\omega_{0}+\delta}$, $z_{2}=\frac{A}{\omega_{0}-\delta}$ and $\gamma=2\delta t+\pi$. 
In the following we use the well-known formulas of summing the Bessel functions for the further transformation of the Hamiltonian. The result reads
\begin{eqnarray}
e^{2i\varphi}= e^{i(\Delta_{N}+N\delta)t-iN\pi}J_{N}(w(t))\left(\frac{z_{2}-z_{1}e^{-i\gamma}}{z_{2}-z_{1}e^{i\gamma}}\right)^{\frac{N}{2}},
\end{eqnarray}
where $w(t)=\left(z_{1}^{2}+z_{2}^{2}-2z_{1}z_{2}\cos(\gamma)\right)^{1/2}$, $|z_{1}e^{\pm i\gamma}|<z_{2}$.
We rewrite the exponent in the following form 
\begin{equation}
e^{\pm2i\varphi(t)}=J_{N}(w(t))e^{\pm i\alpha(t)},
\end{equation}
introducing the function
\begin{equation}\label{alfa}
\alpha(t)=(\Delta_{N}+N\delta)t-N\pi-\frac{iN}{2}\ln\left[\frac{z_{2}-z_{1}e^{-i\gamma}}{z_{2}-z_{1}e^{i\gamma}}\right]. 
\end{equation}
Equation (\ref{alfa}) can be simplified easily if $\delta\ll\omega_{0}$. Indeed, in this case $z_{1}\approx z_{2}$ when $\delta\ll\omega_{0}$, and we can check that the logarithm in Eq. (\ref{alfa}) is simplified as
\begin{eqnarray}
\ln\left[\frac{z_{2}-z_{1}e^{-i\gamma}}{z_{2}-z_{1}e^{i\gamma}}\right]\approx\ln\left(\frac{1-e^{-i\gamma}}{1-e^{i\gamma}}\right)\nonumber\\
=\ln\left[\frac{e^{-\frac{i\gamma}{2}}(e^{\frac{i\gamma}{2}}-e^{-\frac{i\gamma}{2}})}{e^{\frac{i\gamma}{2}}(e^{\frac{-i\gamma}{2}}-e^{\frac{i\gamma}{2}})}\right]=\ln\left(-e^{-i\gamma}\right)\nonumber\\
=i(\pi-\gamma)=i(\pi-2\delta t-\pi)=-2i\delta t.
\end{eqnarray}
Then, in the lowest approximation of $\delta/\omega_{0}$ we obtain $\alpha(t)=\Delta_{N}t-N\pi$ and
\begin{equation}
w(t)\approx 2\frac{A}{\omega_{0}}\left|\cos(\delta t)\right|.
\end{equation}
In this approximation and for the case of exact resonance, $\Delta_{N}=0$, the interaction Hamiltonian is written in the following form
\begin{equation}\label{HamilFine}
\hat{H}_{I}(t)=(-1)^{N+1}\frac{\Delta}{2}J_{N}(w(t))\sigma_{x}.
\end{equation} 

This Hamiltonian, describing the effects of time modulation on qubit dynamics, is nonstationary and $T$ periodic, $H_{I}(t+T)=H_{I}(t)$, with the period $T=\pi/\delta$; thus, QESs and quasienergies can be introduced in this representation. The Hamiltonian is derived for the general case that involves one-quantum resonance process $N=1$ as well as high-order processes with $N>1$. Below we concentrate on consideration of two cases, $N=1$ and $2$, in detail.

\section{Amplitudes of the tunneling and QES}

The different regimes of qubit dynamics in the presence of a time-modulated field are formulated in the adiabatic and diabatic bases in analogy to the case of a monochromatic field \cite{Shev,Hanggi}. The diabatic basis states $\sdown$ and $\sup$ are the eigenstates of the Hamiltonian Eq. (\ref{Ham}), if $\Delta$ and $f(t)$ have vanished. 

\begin{figure}[h]
\begin{math}
 \begin{array}{rr}
    \includegraphics[height=7cm]{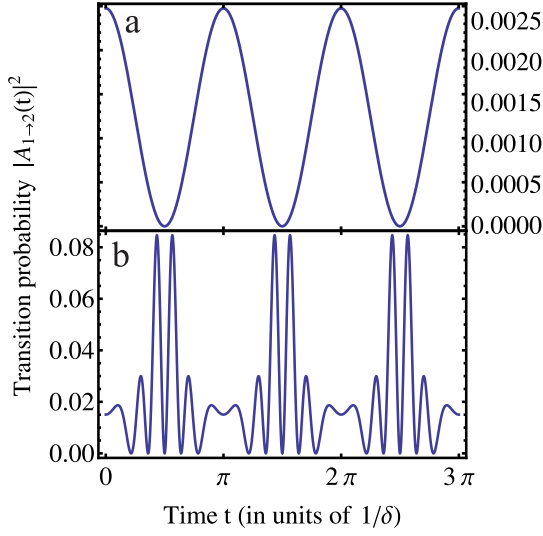}
 \end{array}
\end{math}
  \caption{\label{TransProb1} Transition probabilities for first-order ($N=1$) resonance. The parameters are: (a) $\Delta/\delta=34$, $A/\omega_{0}=10^{-1}$, (b) $\Delta/\delta=34$, $A/\omega_{0}=4.5$.}
\end{figure}

\begin{figure}[h]
\begin{math}
 \begin{array}{rr}
    \includegraphics[height=7cm]{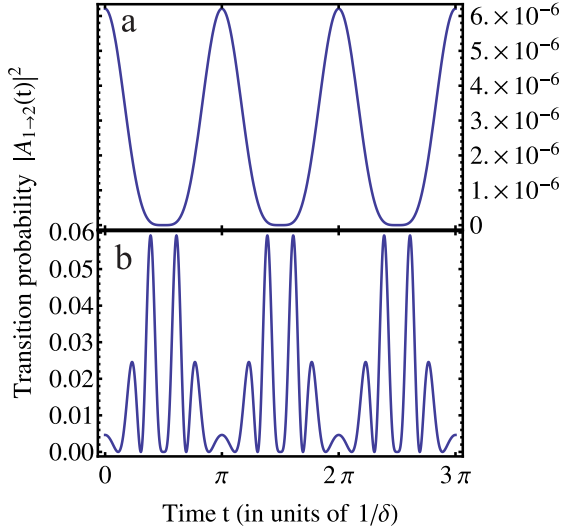}
 \end{array}
\end{math}
  \caption{\label{TransProb2} Transition probabilities for second-order resonance ($N=2$). The parameters are: (a) $\Delta/\delta=34$, $A/\omega_{0}=10^{-1}$; (b) $\Delta/\delta=34$, $A/\omega_{0}=4.5$.}
\end{figure}

Let us consider the case $\varepsilon\gg\Delta$. We assume that states of a qubit are formed in the presence of a driving field and the tunneling process is described by transitions between these states. Then, in the lowest order of the perturbation theory on the basis of Eqs. (\ref{Shred}) and (\ref{HamInt}) the tunneling amplitude in the transition $\sdown\rightarrow\sup$ reads $A_{1\rightarrow 2}=\csup \hat{H}_{I}(t)\sdown=\langle\hspace{0.5mm}\uparrow\hspace{-0.5mm}(t)|\hat{H}_{V}(t)|\hspace{-0.5mm}\downarrow(t)\rangle$, where $|\hspace{-0.5mm}\uparrow(t)\rangle=\hat{U}(t)\sup$, $|\hspace{-0.5mm}\downarrow(t)\rangle=\hat{U}(t)\sdown$ are the diabatic states in the $\hat{U}$ representation. In the limit of a weak driving we have $|\hspace{-0.5mm}\uparrow(t)\rangle=e^{-i\varepsilon t/2}\sup$ and $|\hspace{-0.5mm}\downarrow(t)\rangle=e^{i\varepsilon t/2}\sdown$. For the amplitude we obtain
\begin{equation}\label{amplitud}
A_{1\rightarrow 2}=(-1)^{N+1}\frac{\Delta}{2}J_{N}(w(t)).
\end{equation}
This amplitude describes the tunneling transition in the presence of a time-modulated external field that shifts the energetic levels. It is interesting to compare this result with the analogous one for the case of an external monochromatic field. It is known that in the latter case the amplitude of the transition $\sdown\rightarrow\sup$ with parameters satisfying the resonance does not depend on time intervals, while the amplitude Eq. (\ref{amplitud}) contains time-dependent periodic oscillations at the modulation frequency. In Fig. \ref{TransProb1} and \ref{TransProb2} we depict the corresponding probabilities of the tunneling transition in dependence on dimensionless time for two resonant conditions: $N=1$ and $2$. As we see, the transition amplitudes are not constants and are periodic in time, while for the case of a one-monochromatic driving field these quantities have constant values.

In the case of a weak driving field $A\ll\varepsilon_{0}$ and $\omega_{0}\sim\varepsilon_{0}$, we can use the following approximation for the Bessel function: $J_{n}(x)\sim\frac{x^{n}}{2^{n}n!}$, $x\ll1$; therefore,
\begin{equation}\label{BessApprox}
J_{N}(w(\tau))=\frac{1}{N!}\left(\frac{A}{\omega_{0}}\right)^{N}|\cos(\delta t)|^{N}.
\end{equation}
Thus, the amplitude of tunneling for a weak driving field is calculated as 
\begin{equation}\label{amplitudW}
A_{1\rightarrow 2}=(-1)^{N+1}\frac{\Delta}{2}\frac{1}{N!}\left(\frac{A}{\omega_{0}}\right)^{N}|\cos(\delta t)|^{N}.
\end{equation}
This result is in accordance with the results of numerical calculations  corresponding to first-order and second-order resonances presented in Fig.\ref{TransProb1}(a) and \ref{TransProb2}(a). 

Below we turn to the general case of qubit dynamics considering the state of the full system in the $\sdown$, $\sup$ basis as
\begin{equation}
|\Psi_{U}(t)\rangle=C_{1}(t)\sdown+C_{2}(t)\sup.
\end{equation} 

In this case, the Schr{\"o}dinger equation is reduced to two coupled first-order equations for the amplitudes in the following form
\begin{subequations}\label{c1c2eq}
\begin{eqnarray}
i\dot{C}_{1}(t) & = & -\frac{\Delta}{2}J_{N}(w(t))e^{-i\alpha(t)}C_{2}(t),\\
i\dot{C}_{2}(t) & = & -\frac{\Delta}{2}J_{N}(w(t))e^{i\alpha(t)}C_{1}(t).
\end{eqnarray}
\end{subequations}
The coefficients of these equations have a nontrivial dependence on time, nevertheless we demonstrate that for the resonance case, $\Delta_{N}=0$, the solution of these equations can be found in a simple analytical form as follows:
\begin{subequations}\label{c1c2}
\begin{eqnarray}
C_{1}(t) & = & \cos(\gamma_{N}(t)),\\
C_{2}(t) & = & i e^{i\alpha}\sin(\gamma_{N}(t)),
\end{eqnarray}
\end{subequations}
while the function $\gamma_{N}(t)$ is calculated from Eqs. (\ref{c1c2eq}a) and (\ref{c1c2eq}b) as
\begin{equation}\label{GammaInt}
\gamma_{N}(t)=\frac{\Delta}{2}\int_{0}^{t}J_{N}(w(\tau))d\tau
\end{equation}
and $\alpha(t)=-N\pi$. 

This solution is presented for the concrete initial conditions assuming that the system is initially in the lower state; therefore, $C_1(0)=1$ and $C_2(0)=0$. The populations of the initial and excited states (if the system was initially in the lower state) as a function of time are then given by
\begin{subequations}\label{pop}
\begin{eqnarray}
P_{1}(t)=|C_{1}(t)|^{2}=\cos^{2}(\gamma_{N}(t)),\\
P_{2}(t)=|C_{2}(t)|^{2}=\sin^{2}(\gamma_{N}(t)).
\end{eqnarray}
\end{subequations}
To calculate these quantities further we need to analyze the function $\gamma_{N}(t)$ that involves integration of a periodic function $J_{N}(w(\tau))$ with period $T=\pi/\delta$. It is easy to represent the function $\gamma_{N}(t)$ as
\begin{equation}\label{GammaPeriodic}
\gamma_{N}(t)=\frac{\Delta}{2}\overline{J_{N}}t+\Phi_{N}(t),
\end{equation}
where 
\begin{equation}\label{Jmean}
\overline{J_{N}}\equiv\overline{J_{N}(w(t))}=\frac{1}{T}\int^{t_{0}+T}_{t_{0}}J_{N}(w(\tau))d\tau,
\end{equation}
$\Phi(t)$ is a periodic function defined for $t\in[t_{0},t_{0}+T]$ as
\begin{equation}\label{fi}
\Phi_{N}(t)=\frac{\Delta}{2}\int^{t}_{t_{0}}(J_{N}(w(\tau))-\overline{J_{N}})d\tau,
\end{equation}
and for other $t\in[0,\infty]$ through periodicity relation $\Phi(t+T)=\Phi(t)$ (see the Appendix). 

The above formulas allow us to introduce the QES of a qubit in a time-modulated driving field. Indeed, it is easy to check that the solution of Eq. (\ref{Shred}) with periodic in time Hamiltonian (\ref{HamilFine}) can be expressed in the adiabatic basis as    
\begin{equation}
|\Theta_{N,\pm}(t)\rangle=e^{\pm i(-1)^{N}\gamma_{N}(t)}|\varphi_{\pm}\rangle,
\end{equation}
where 
\begin{equation}
|\varphi_{\pm}\rangle=\sdown\pm\sup.
\end{equation}
Then, by using the formula Eq. (\ref{GammaPeriodic}), these states can be presented in the form of a QES
\begin{equation}\label{CircuitState}
|\Theta_{N,\pm}(t)\rangle=e^{iE^{\pm}_{N}t}U_{N,\pm}(t)|\varphi_{\pm}\rangle,
\end{equation}
where
\begin{equation}
U_{N,\pm}(t)=e^{\pm i(-1)^{N}\Phi_{N}(t)}
\end{equation}
are periodic in time; $U_{N,\pm}(t+T)=U_{N,\pm}(t)$, and $E^{\pm}_{N}=\pm E_{N}$, where 
\begin{equation}\label{QuasiE}
E_{N}=(-1)^{N}\frac{\Delta}{2}\overline{J_{N}} 
\end{equation}
are the quasienergies. In the $\Psi$ representation we obtain
\begin{equation}
|\Psi_{\pm,N}\rangle=e^{iE^{\pm}_{N}t}U_{N,\pm}(t)\left(e^{-i\varphi(t)}\sdown\pm e^{i\varphi(t)}\sup\right).
\end{equation}

\section{Quasienergies of the qubit in bichromatic field}

In this seciton we study properties of the quasienergies. Note that some experiments recently realized in the field of superconducting Josephson qubits have been interpreted in terms of the probe absorption spectroscopy of the quasienergy levels (see, for example, \cite{Silveri}). In this way, the frequencies of probe field absorption or amplification are determined by the matrix elements of transition between QESs. We briefly discuss this problem, considering the transition $|\Theta_{N,+}\rangle\rightarrow|\Theta_{N,-}\rangle$ between quasienergetic states $\Theta_{N,\pm}(t)$ due to a weak interaction of the system with a probe field. Such an interaction with a probe field at the frequency $\omega_{p}$ can be added as weak perturbation term $\lambda E_{p}\cos(\omega_{p}t)\sigma_{z}$ in the Hamiltonian (\ref{Ham}).  Thus, the matrix element of this transition is calculated as
\begin{eqnarray}
\langle\Theta_{N,-}|\sigma_{z}|\Theta_{N ,+}\rangle\nonumber\\
=2\sum_{n,m}J_{n}\left(\frac{A}{\omega_{0}}\right)J_{m-n}\left(\frac{A}{\omega_{0}}\right)\exp(i\Omega_{m,n}t),
\end{eqnarray}
where $m=0,\pm1,\pm2,...$; $n=0,\pm1,\pm2, ...$; and $\Omega_{m,n}=\varepsilon_{0}+m\omega_{0}+(E^{+}-E^{-})+(2n-m)\delta$ are the frequencies of spectral lines corresponding to the absorption (for $\omega_{p}=\Omega_{m,n}>0$) and the amplification (for $ \omega_{p}=\Omega_{m,n}<0$) of a probe field. As we can see, the spectral lines separated by the central frequency and modulation harmonics and contain a field-dependent Stark shift due to the input of the quasienergies. In Sec. V we demonstrate that the quasienergies in Eq. (\ref{QuasiE}) also play an essential role in occupation populations of states.

As we see, the sum of two quasienergies obeys the relation $E^{+}_{N}+E^{-}_{N}=0$. This result is in accordance with the exact result taking place for a two-level atom in a monochromatic field. According to \cite{Shirley} the sum of two quasienergies equals the sum of atomic energetic levels, that is zero for the case of the truncated Hamiltonian Eq. (1), in which the half of the sum of qubit energetic levels has been omitted.

\begin{figure}[h]
 \begin{math}
 \begin{array}{cc}
    \includegraphics[height=4.8cm]{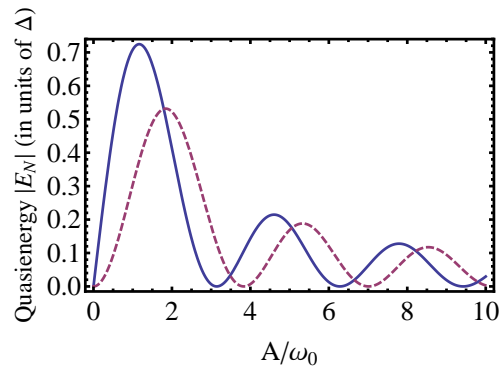}
 \end{array}
\end{math}
  \caption{\label{QuasiEnergy} Quasienergy for ($N=1$) first-order resonance (solid curve); ($N=2$) second-order resonance (dashed curve).}
\end{figure}

The difference between quasienergies reads $E^{+}_{N}-E^{-}_{N}=2E_{N}$ in this case. In this representation quasienergies contain only a field-dependent part and equal zero in the limit of small driving. Dependences of the quasienergy $E^{\pm}_{N}$ on the parameter $A/\omega_{0}$ as a function $E^{\pm}_{N}=E^{\pm}_{N}(A/\omega_{0})$ for two types of resonances, for the first order as well as for the second order, are shown in Fig.\ref{QuasiEnergy}.
As we demonstrate, the quasienergies $E_{1}$ and $E_{2}$ for both types of resonances have zeros for the definite values of $A/\omega_{0}$. The lower zeros are at $A/\omega_{0}=3.13, 6.3,$ and $9.45$ for $N=1$ and are at $A/\omega_{0}=3.8, 7.05,$ and $10.2$ for $N=2$. Below, we also analyze the quasienergies for the regime of a weak external field.

\subsection{Quasienergies and phase function at the regime of weak driving}

In this subsection we derive approximative analytical results for the quasienergies and the phase function Eq. (\ref{GammaPeriodic}) using the formula Eq. (\ref{BessApprox}), which describes the weak driving limit. Integration of the formula (\ref{Jmean}) leads to  
\begin{eqnarray}\label{JmeanApp}
\overline{J_{N}}=\frac{\delta}{\pi}\int^{\pi/\delta}_{0}J_{N}(w(\tau))d\tau=\frac{F_{[0;\pi],N}(\pi/\delta)}{N!\pi}\left(\frac{A}{\omega_{0}}\right)^{N}\nonumber\\
=\frac{2\Gamma\left(\frac{3+N}{2}\right)+(1+N)\Gamma\left(\frac{1+N}{2}\right)}{2\sqrt{\pi}N!(1+N)\Gamma\left(\frac{3+N}{2}\right)}\left(\frac{A}{\omega_{0}}\right)^{N}.
\end{eqnarray}
In this formula, we introduce a function, which also determines the periodic part of the phase function Eq. (\ref{fi}),
\begin{eqnarray}
F_{[0;\pi],N}(t)=\frac{\sqrt{\pi}}{2}\frac{\Gamma\left(\frac{1+N}{2}\right)}{\Gamma\left(1+\frac{N}{2}\right)}\nonumber\\
-\frac{\prescript{}{2}F_{1}\left(\frac{1}{2},\frac{1+N}{2};\frac{3+N}{2};\cos^{2}(\delta t)\right)}{1+N}\cos(\delta t)|\cos(\delta t)|^{N}
\end{eqnarray}
which is defined in $[0;\pi/\delta]$. Here, $\prescript{}{2}F_{1}(a,b;c;z)$ is a hypergeometric function. The final result for the quasienergy reads as follows: 
\begin{eqnarray}
{E_{N}}=
(-1)^{N}\frac{\Delta}{2}\frac{2\Gamma\left(\frac{3+N}{2}\right)+(1+N)\Gamma\left(\frac{1+N}{2}\right)}{2\sqrt{\pi}N!(1+N)\Gamma\left(\frac{3+N}{2}\right)}\left(\frac{A}{\omega_{0}}\right)^{N},
\end{eqnarray}
while the periodic part of the phase function is calculated as
\begin{eqnarray}\label{fiApp}
\Phi_{N}(t)=\frac{\Delta}{2}\int^{t}_{0}(J_{N}(w(\tau))-\overline{J_{N}})d\tau\nonumber\\
=\frac{\Delta}{\delta}\frac{1}{2N!}\left(\frac{A}{\omega_{0}}\right)^{N}[F_{[0;\pi],N}(t)-F_{[0;\pi],N}(\pi/\delta)\delta t].
\end{eqnarray}

Note, that the results of this section on the quasienergetic states of the qubit in a time-modulated driving field are essentially different from the analogous results for the states obtained for a two-level atomic system driven by a bichromatic field with the Hamiltonian Eq. (\ref{AtomHam}) \cite{Kryuchkov,Freed1,Jakob1,Jakob2,Jakob3}. We demonstrate this point below.

\subsection{System with time-dependent component along $x$ axis}

In this subsection we briefly discuss the system with the Hamiltonian Eq. (\ref{AtomHam}), which is typical for problems that involve an atom in a bichromatic laser field. Our goal is to show the differences of the behaviors for the cases of superconducting qubits [see Hamiltonian Eq.(\ref{Ham} with the time-dependent component along $z$axis] and two-level atomic system [see Hamiltonian Eq.(\ref{AtomHam}) with the time-dependent component along $x$ axis) in bichromatic field.

We now take the system described by the Hamiltonian Eq. (\ref{AtomHam}) in new denotations that are more standard in this area:  
\begin{equation}\label{AtomHam2}
\hat{H}=-\frac{\Delta E}{2}\hat{\sigma}_{z}+V\cos(\omega_{0}t)\cos(\delta t)\hat{\sigma}_{x}.
\end{equation}
We make a transformation to a rotating frame $|\Psi(t)\rangle=\hat{W}(t)|\Psi_{W}(t)\rangle$, where
\begin{equation}
\hat{W}(t)=\exp\left(\frac{i}{2}\Delta E t\hat{\sigma}_{z}\right).
\end{equation}
For this system we can formulate only a one-quantum condition of the resonance, $\Delta E=\omega_{0}$, in contrast to the system with the time-dependent component along the $z$ axis in which multiquantum resonances take place.  In the resonance approximation we obtain 
\begin{equation}
i\frac{\partial}{\partial t}|\Psi_{W}(t)\rangle=V\cos(\delta t)\hat{\sigma}_{x}|\Psi_{W}(t)\rangle.
\end{equation}
The solution of this equation in the adiabatic bases can be obtained as
\begin{equation}\label{AtomState}
|\Phi_{\pm}(t)\rangle=\exp\left(i\frac{V}{\delta}\sin(\delta t)\right)|\varphi_{\pm}\rangle.
\end{equation}

Comparing the results of Eqs. (\ref{CircuitState}) and (\ref{AtomState}) we conclude that the quasienergies  corresponding  to QES Eq. (\ref{AtomState})  are equal to zero for all ranges of the parameters in the rotating wave approximation in contrast to the results of Eqs. (\ref{CircuitState}), (\ref{QuasiE}). Besides this, the periodic wave function $\exp\left[i\frac{V}{\delta}\sin(\delta t)\right]$ strongly differs from the periodic wave function $U_{N,\pm}(t)$ that corresponds to QES Eq. (\ref{CircuitState}). This situation is displayed also in the  frequencies of  spectral lines corresponding to the transitions between QES $|\Phi_{\pm}(t)\rangle$. Indeed, it is easy to realize that these frequencies are at $\omega_{p}=\omega_{0}+n\delta$, $n=0,\pm1,\pm2,...$ and do not involve field-dependent shifts of energetic levels. This effect is in accordance with calculation of the spectrum of resonance fluorescence and Autler-Townes splitting in a bichromatic field \cite{Jakob2,Jakob3}.

At the end of this section, for completeness, we present the QES in the $\Psi$ representation: 

\begin{eqnarray}
|\Psi_{+}(t)\rangle=\hat{W}(t)|\Phi_{+}(t)\rangle\nonumber\\
=\frac{1}{\sqrt{2}}\exp\left(i\frac{V}{\delta}\sin(\delta t)\right)\left(e^{-i\frac{\Delta E t}{2}}\sdown+e^{i\frac{\Delta E t}{2}}\sup\right),\\
|\Psi_{-}(t)\rangle=\hat{W}(t)|\Phi_{-}(t)\rangle\nonumber\\
=\frac{1}{\sqrt{2}}\exp\left(-i\frac{V}{\delta}\sin(\delta t)\right)\left(e^{-i\frac{\Delta E t}{2}}\sdown-e^{i\frac{\Delta E t}{2}}\sup\right).
\end{eqnarray}

It should be noted that the Floquet basis derived here for a qubit in a bichromatic field is useful for studying the Rabi oscillation physics as well as for writing the master equation governing the dynamics of the reduced density matrix of a driven system, which is in contact with an external environment.

\begin{figure}[h]
    \includegraphics[height=4cm]{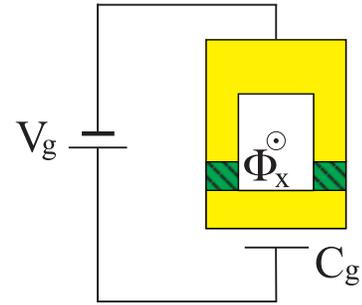}
\caption{\label{circuit} A charge qubit with tunable effective Josephson coupling. It is controlled by $V_{g}$ gate voltage and $\Phi_{x}$ magnetic field.}
\end{figure}

Note that Hamiltonians Eq. (\ref{Ham}) and (\ref{AtomHam2}) are the particular cases of a more general Hamiltonian:
\begin{equation}
\hat{H}(t)=-\frac{1}{2}B_{z}(t)\hat{\sigma}_{z}-\frac{1}{2}B_{x}(t)\hat{\sigma}_{x},
\end{equation}
which can be realized on a properly designed superconducting circuit. In particular, a simple design of the charge qubit with tunable effective Josephson coupling can be shown schematically (see Fig.\ref{circuit}) as
with time-dependent coefficients  $B_{x}(t)$ and $B_{z}(t)$ that allow complete control of the system through the gate voltage $V_{g}$ and the external magnetic flux $\Phi_{x}$ (see, for example \cite{Makh}). The relations between these quantities are expressed as
\begin{eqnarray}
B_{x}(t)=E_{J}(\Phi_{x}(t))=2E_{J}^{0}\cos\left(\pi\frac{\Phi_{x}(t)}{\Phi_{0}}\right),\\
B_{z}(t)=\delta E_{ch}(V_{g}(t))=4E_{C}\left(1-\frac{1}{e}C_{g}V_{g}(t)\right),
\end{eqnarray}
where $E_{C}$ is the the single-electron charging energy, $E_{J}^{0}$ is the Josephson coupling energy, and $C_{g}$ is the gate capacitor. Practical realizations of an analogous scheme have been done in a series of papers (see, for example, \cite{Oliver,Berns}). We believe that the schemes considered here with a time-dependent component along the $z$ or $x$ axis driven by bichromatic external fields can be constructed in the same way (see \cite{Paraoanu}, in which multisideband components of qubit energy splitting are observed).

\section{Aperiodic and periodic Rabi oscillations}

\begin{figure}[h]
 \begin{math}
 \begin{array}{cc}
    \includegraphics[height=4.5cm]{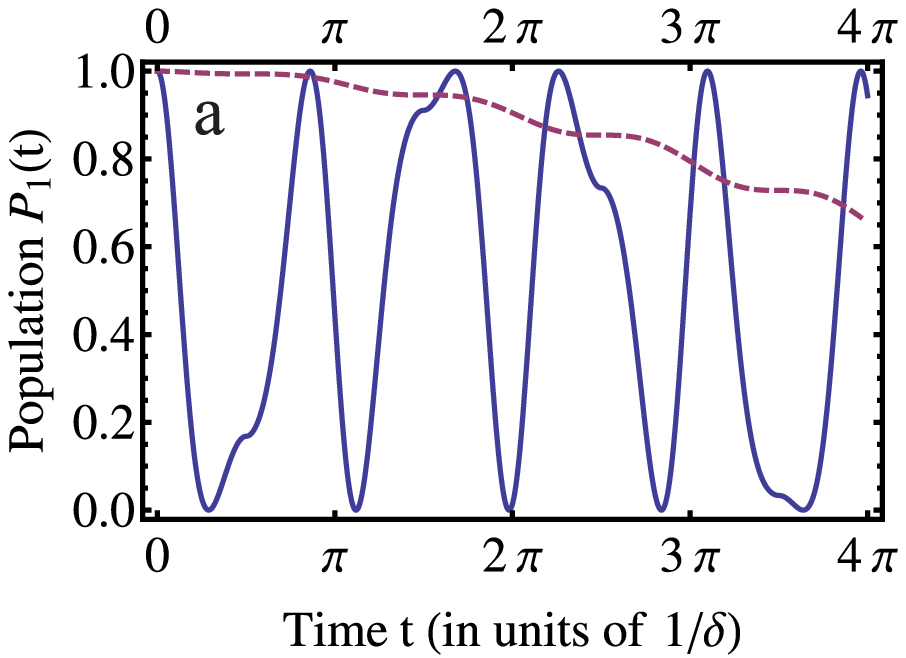}\\
    \includegraphics[height=4.5cm]{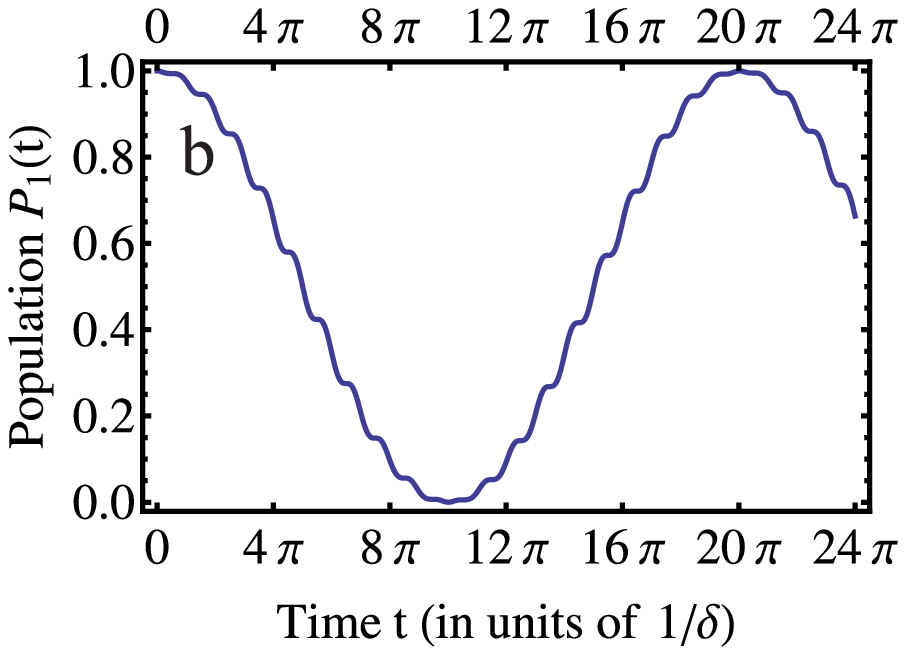}\\
    \includegraphics[height=4.5cm]{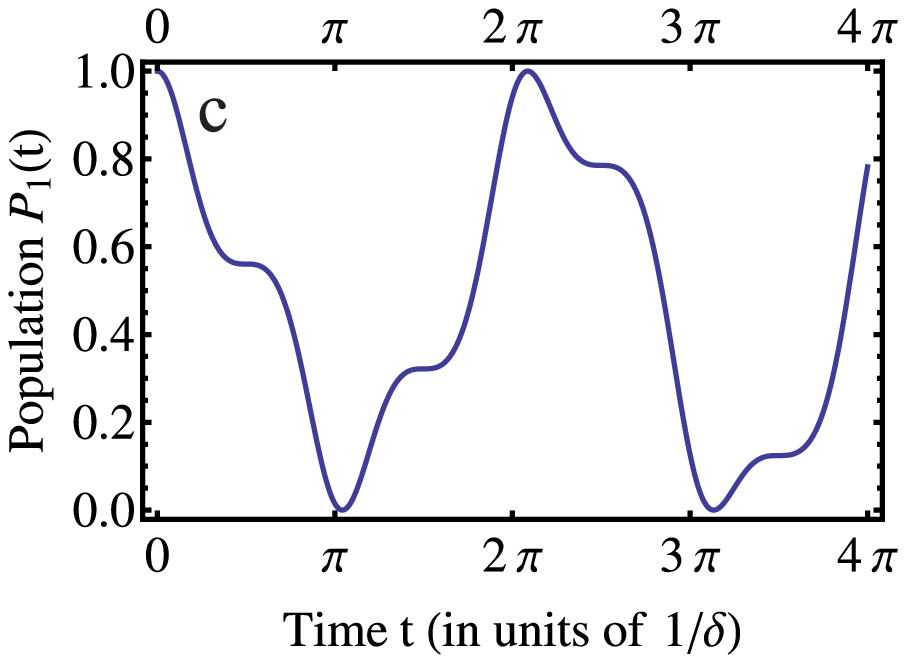}
 \end{array}
\end{math}
  \caption{\label{pop3} Populations of the ground state for both resonance conditions: ($N=1$) and ($N=2$). The parameters are: (a) first (solid curve) and second (dashed curve) order resonances, $\Delta/\delta=40$, $A/\omega_{0}=10^{-1}$; (b) second order resonance ($N=2$) for the same values of $\Delta/\delta$ and $A/\omega_{0}$ as in (a); (c) second order resonance ($N=2$), $\Delta/\delta=370$, $A/\omega_{0}=10^{-1}$.}
\end{figure}

\begin{figure}[h]
 \begin{math}
 \begin{array}{cc}
    \includegraphics[height=7cm]{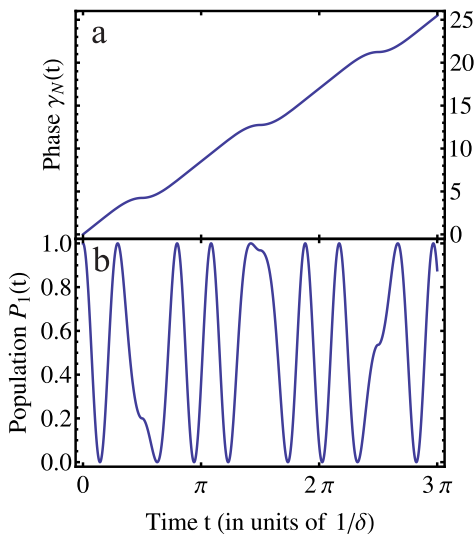}
 \end{array}
\end{math}
  \caption{\label{Population1} (a) Phase-function $\gamma_{N}(t)$ and (b) population probability for the first-order resonance condition ($N=1$). The parameters are:   $\Delta/\delta=12$, $A/\omega_{0}=1$.}
\end{figure}

\begin{figure}[h]
 \begin{math}
 \begin{array}{cc}
    \includegraphics[height=7cm]{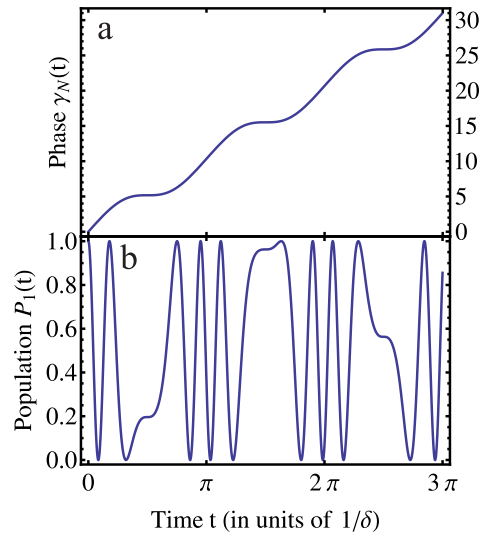}
 \end{array}
\end{math}
  \caption{\label{Population2} (a) Phase-function $\gamma_{N}(t)$  and (b) population probability for the second-order resonance condition ($N=2$). The parameters are:  $\Delta/\delta=34$, $A/\omega_{0}=1$.}
\end{figure}

In this section, time-dependent populations of states are investigated for various regimes. We investigate dynamics of the driven qubit in a time domain for various resonance conditions. Thus, we consider the occupation probability as a function of time in dimensionless units, assuming that the system was initially in the state $\sdown$ and the Rabi frequency is given by
\begin{equation}
\Omega_{N}(t)=\dot{\gamma}_{N}(t).
\end{equation}
According to the formulas Eqs. (\ref{c1c2}) and (\ref{pop}) this dynamics is determined by the function $\gamma_{N}(t)$, that involves both the quasienergy and the periodic function $\Phi_{N}(t)$ with period $T=\pi/\delta$. To present this statement in a clear form we rewrite the formula Eq. (\ref{GammaPeriodic}) as  
\begin{equation}\label{GammaQuasienerg}
\gamma_{N}(t)=(-1)^{N}E_{N}t+\Phi_{N}(t).
\end{equation}
The function $\gamma_{N}(t)$ is an increasing function in time but it grows also periodically due to its "linear+periodic" structure. Therefore, the dynamics of populations Eq. (\ref{pop}) seems to be aperiodic in time. Indeed, the typical results for the phase function as well as the populations are depicted in Figs.\ref{pop3},\ref{Population1} and \ref{Population2}. The dynamics of populations for the case of a weak external field is shown in Figs.\ref{pop3} for two resonance regimes. In Fig.\ref{pop3}(a) we compare two curves of the occupation probabilities for $N=1$ (solid curve) and for $N=2$ (dashed curve). We can see here fast oscillations of the population for the regime $N=1$ and slow oscillations for the case of $N=2$ (for consideration in details, see the curve corresponding to the case $N=2$ for large time intervals in Fig.\ref{pop3}(b)). The results for the second-order resonance regime are also demonstrated in Fig.\ref{pop3}(c) for the other parameter $\Delta/\delta$. Analyzing these results, we note that dynamics of populations strongly depends on the value of the ratio $\Delta/\delta$. It can be seen from the formulas Eqs. (\ref{JmeanApp}) and (\ref{fiApp}) that population behavior shown in Fig.\ref{pop3}(b) for $N=2$ is mainly governed by the linear in time term in the phase function Eq. (\ref{fiApp}); thus, we can see that the dynamics looks like cosinusoidal oscillations. The periodic in time part $\Phi_{N}(t)$ only slightly modulated these oscillations. This part of the phase function increases with increasing the parameter $\Delta/\delta$ that leads to increasing the role of periodic modulations giving rise to a nontrivial time dependence of occupation probability [see, Fig.\ref{pop3}(c) for the case $N=2$].

The typical examples of occupation probabilities corresponding to the large-amplitude regime are depicted in Figs. \ref{Population1} and \ref{Population2}. In order to illustrate the role of phase function in the development of aperiodic dynamics we also show here the time dependence of this function.

It is obvious that the dynamics of populations can be periodic if the quasienergy becomes equal to zero at definite values of the parameter $A/\omega_{0}$ (see, Fig.\ref{QuasiEnergy}). However, as it can be seen, this situation also takes  place for the other wide ranges of the parameters, if the shift of the phase function during $m$ periods $T=\pi/\delta$ 
\begin{eqnarray}
\gamma(t+mT)=(-1)^{N}E_{N}(t+mT)+\Phi_{N}(t)\nonumber\\
=\gamma(t)+(-1)^{N}\frac{E_{N}\pi}{\delta}m
\end{eqnarray}
becomes equal to $n\pi$, that is the period of square $\cos(x)$ or $\sin(x)$ in the formulas Eq. (\ref{pop}). Such a consideration leads to the following formula:
\begin{equation}\label{ResCond}
m\delta=n|E_{N}|,
\end{equation}
where $m$ and $n$ are positive integers. The physical means of this formula is very simple. The population of the states depends on $\gamma_{N}(t)$ as a square of the cosine, for example $P_{1}=\cos^{2}[\gamma_{N}(t)]$. Thus, if during $m$ periods its growth is equal to any period of $\cos^{2}(x)$, which can be written as $n\pi$, the population will repeat its behavior. 

\begin{figure}[h]
\begin{math}
 \begin{array}{rr}
    \includegraphics[height=4.5cm]{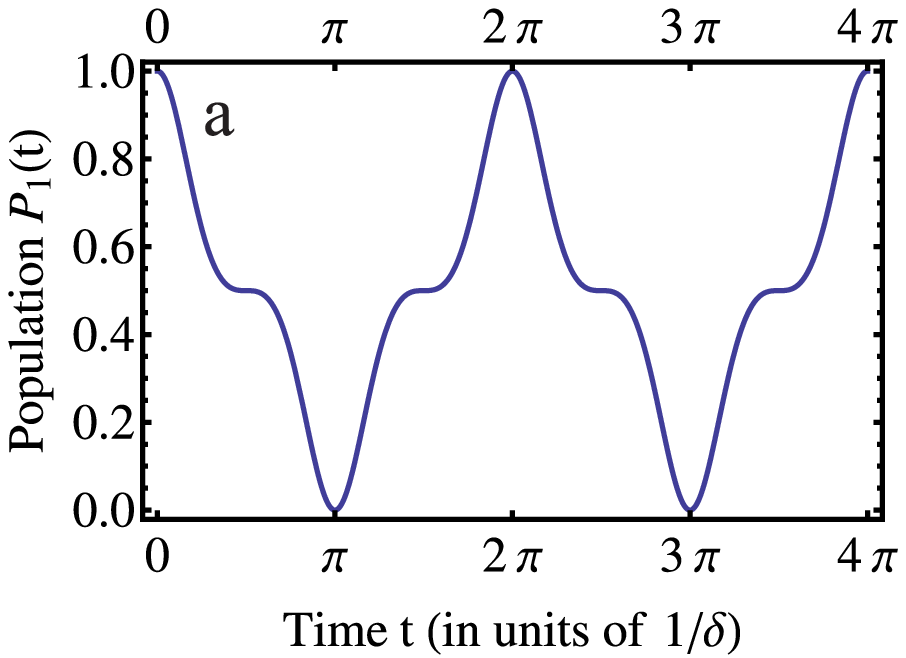}\\
    \includegraphics[height=4.5cm]{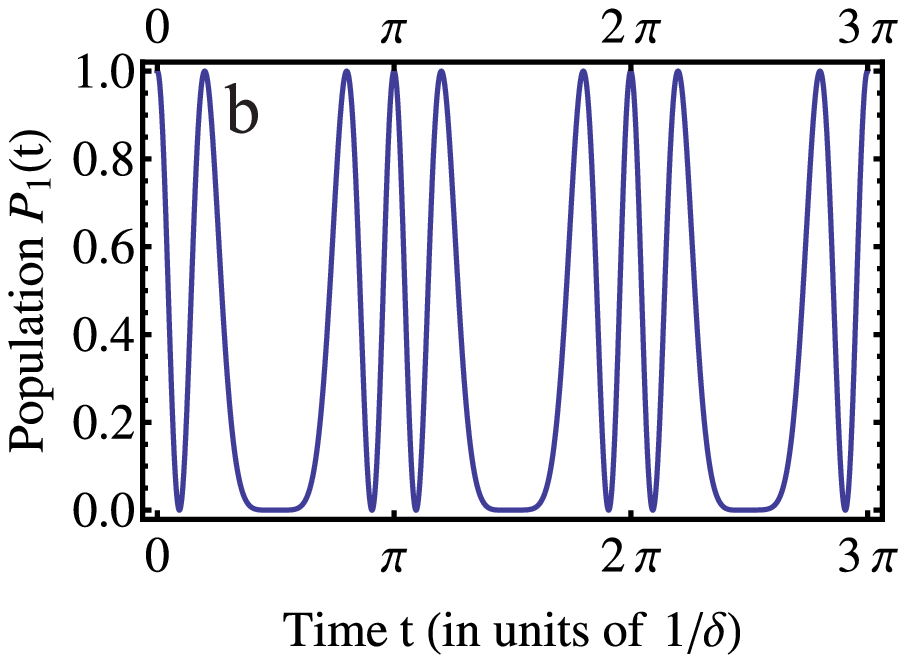}
 \end{array}
\end{math}
  \caption{\label{Periodic} The periodically regular dynamics of population probabilities for the second-order resonance. The parameters are: (a) $\Delta/\delta=401$, $A/\omega_{0}=10^{-1}$;  (b) $\Delta/\delta=31$, $A/\omega_{0}=1$.}
\end{figure}

Thus, the populations could be made periodic by choosing the values of parameters $\delta/\Delta$ and $A/\omega_{0}$ to satisfying the following condition: 
\begin{equation}\label{periodrelation}
\frac{\delta}{\Delta}=\frac{n}{m}\frac{1}{2\pi}\int_{0}^{\pi}J_{N}\bigg(2\frac{A}{\omega_{0}}|\cos(\tau)|\bigg)d\tau,
\end{equation}
that follows from the formulas Eqs. (\ref{QuasiE}) and (\ref{ResCond}). For the case of a weak driving field this condition is simplified and reads 
\begin{equation}
\frac{1}{N!}\frac{\Delta}{2\delta}\left(\frac{A}{\omega_{0}}\right)^{N}*\sqrt{\pi}\frac{\Gamma\left(\frac{1+N}{2}\right)}{\Gamma\left(1+\frac{N}{2}\right)}=\pi\frac{m}{n}.
\end{equation}

The typical results for Rabi oscillations with regular, periodic dynamics are depicted in Fig.\ref{Periodic} for the $N=2$ resonance condition. Here, the parameters $A/\omega_{0}$ and two used parameters, $\Delta/\delta=401$ [see Figs.\ref{Periodic}(a)] and $\Delta/\delta=31$ [see Figs.\ref{Periodic}(b)], satisfy the periodicity condition Eq. (\ref{periodrelation}). We compare the results shown in Fig.\ref{Periodic}(a) with the result depicted in Fig.\ref{pop3}(c). Both results are obtained for the second-order resonance condition and for the same parameter $A/\omega_{0}=10^{-1}$; however, using the parameter $\Delta/\delta$ satisfying the condition of periodicity Eq. (\ref{periodrelation}) in Fig.\ref{Periodic}(a) leads to the periodic dynamics of the populations. These regimes in which quantum dynamics of occupation probabilities becomes periodically regular can be useful, for example, in applications where one is dealing with logic operations on qubits.

\section{Conclusion}
In conclusion, we have analyzed dynamics of a superconducting qubit interacting with an electromagnetic wave with time-modulated amplitude (or a bichromatic field) for two basic configurations that involve time-dependent components along $z$ or $x$ axes. In the case of the $z$ configuration, the external bichromatic field drives the qubit's energetic levels, while in the case of $x$ configuration (describing also the standard problems of a two-level atom in a bichromatic field) the coupling with the bichromatic field leads to the transition dipole moment between two states of atoms. We have calculated quasienergetic states and quasienergies of the composite system "superconducting qubit plus time-modulated field" in an adiabatic basis of the system analyzing the quasienergies  numerically for arbitrary intensities of the external field as well as analytically in detail for the regime of weak driving. Considering the dependence of quasienergies from the intensity parameter $A/\omega_{0}$ we have shown oscillation-type behavior of quasienergies for the case of a strong bichromatic field. In this way, we demonstrate the drastic difference between QESs of two schemes. Particularly, for the standard two-level model in a bichromatic field (the $x$ configuration) the quasienergies are equal to zero for all ranges of the parameters that are displayed in the spectral line of RF and Autler-Townes splitting \cite{Kryuchkov,Freed1,Jakob1,Jakob2,Jakob3}. In contrast to this case, the QES for the scheme involving time-dependent $z$-axis coupling has a more complicated structure. On the whole, the spectral lines of QES transitions  contain also field-dependent Stark shifts due to the input of the quasienergies.

We have considered time dependence of the occupation probabilities of qubit states and Rabi physics for both first-order ($N=1$) and second-order ($N=2$) resonance regimes, when the central frequency $\omega_{0}$ and the electronic energy difference obey rules $\varepsilon_{0}=N\omega_{0}$. Considering Rabi oscillations between qubit states we have shown that these oscillations are aperiodic in time due to effects of time-dependent modulation. Nevertheless, further, we have demonstrated new regimes in which dynamics of populations becomes periodically regular. These regimes can be realized if the ratio of quasienergy to the detuning is positive integer $E_{N}/\delta=r$ for an arbitrary order of resonances. Together with the recent advancements in the engineering of various schemes of superconducting qubits, these results seem to be important for further studies of quantum phenomena in this area.

\appendix*
\section{}
The formula Eq. (\ref{GammaPeriodic}) can be derived by using the Fourier expansion
\begin{equation}
J_{N}(w(t))=\sum^{\infty}_{n=-\infty}G(n)e^{i \frac{2\pi n}{T}t}.
\end{equation}
Then, the integral from Eq. (\ref{GammaInt}) is transformed to
\begin{eqnarray}\label{intderiv}
\int_{0}^{t}J_{N}(w(\tau))d\tau=\sum^{\infty}_{-\infty}G(n)\int^{t}_{0}e^{-i \frac{2\pi n}{T}t'}dt'\nonumber\\
=\sum^{\infty}_{|n|=1}G(n)\frac{T}{i 2\pi n}(e^{i \frac{2\pi n}{T}t}-1)+G(0)t.
\end{eqnarray}
Here, it is easy to realize that 
\begin{equation}
G_{N}(0)=\overline{J_{N}(w(t))}=\overline{J_{N}},
\end{equation}
and thus the formula Eq. (\ref{GammaPeriodic}) is obtained. In this representation the second term of the function Eq. (\ref{GammaPeriodic}) is written in the following form:  
\begin{equation}
\Phi_{N}(t)=\frac{\Delta}{2}\sum^{\infty}_{|n|=1}G(n)\frac{T}{i 2\pi n}(e^{i \frac{2\pi n}{T}t}-1).
\end{equation}

\nocite{*}

\bibliographystyle{apsrev4-1}
\bibliography{abovyan}% Produces the bibliography via BibTeX.

\end{document}